\documentclass[a4paper,twocolumn,superscriptaddress,11pt,unpublished]{quantumarticle}
\pdfoutput=1
\usepackage[utf8]{inputenc}
\usepackage[english]{babel}
\usepackage[T1]{fontenc}
\usepackage{amsmath}
\usepackage[colorlinks,pdftex]{hyperref}

\usepackage{tikz}
\usepackage{lipsum}

\usepackage{amsthm}
\usepackage{graphicx}
\usepackage{epstopdf}
\usepackage{bm}
\usepackage{amsmath,amssymb,amsfonts}
\usepackage{xcolor}

\usepackage{hyperref}
\usepackage{braket}

\usepackage[sort&compress,numbers]{natbib}

\usepackage[normalem]{ulem}
\newcommand\erase{\bgroup\markoverwith{\textcolor{red}{\rule[.5ex]{2pt}{0.4pt}}}\ULon}
\newcommand\eraseblue{\bgroup\markoverwith{\textcolor{blue}{\rule[.5ex]{2pt}{0.8pt}}}\ULon}
\begin{document}

\title{Light-cone feature selection for quantum machine learning 
}
\date{}
\author{Yudai Suzuki*}
\affiliation{Department of Mechanical Engineering, Keio University, Hiyoshi 3‑14‑1, Kohoku, Yokohama, 223‑8522, Japan}
\author{Rei Sakuma}
\affiliation{Materials Informatics Initiative, RD Technology \& Digital Transformation Center, JSR Corporation, 3-103-9, Tonomachi, Kawasaki-ku,
Kawasaki, Kanagawa, 210-0821, Japan}
\affiliation{Quantum Computing Center, Keio University, Hiyoshi 3‑14‑1, Kohoku, Yokohama, 223‑8522, Japan}
\author{Hideaki Kawaguchi}
\affiliation{Quantum Computing Center, Keio University, Hiyoshi 3‑14‑1, Kohoku, Yokohama, 223‑8522, Japan}


\maketitle

\begin{abstract}
Feature selection plays an essential role in improving the predictive performance and interpretability of trained models in classical machine learning. 
On the other hand, the usability of conventional feature selection could be limited for quantum machine learning tasks; the technique might not provide a clear interpretation on embedding quantum circuits for classical data tasks and, more importantly, is not applicable to quantum data tasks. 
In this work, we propose a feature selection method with a specific focus on quantum machine learning.
Our scheme treats the light-cones (i.e., subspace) of quantum models as features and then select relevant ones through training of the corresponding local quantum kernels.
We numerically demonstrate its versatility for four different applications using toy tasks: (1) feature selection of classical inputs, (2) circuit architecture search for data embedding, (3) compression of quantum machine learning models and (4) subspace selection for quantum data. 
The proposed framework paves the way towards applications of quantum machine learning to practical tasks. Also, this technique could be used to practically test if the quantum machine learning tasks really need quantumness, while it is beyond the scope of this work.
\end{abstract}

\section{Introduction}
\label{intro}

\begin{figure*}[htp]
  \begin{center}
    \includegraphics[scale=0.8]{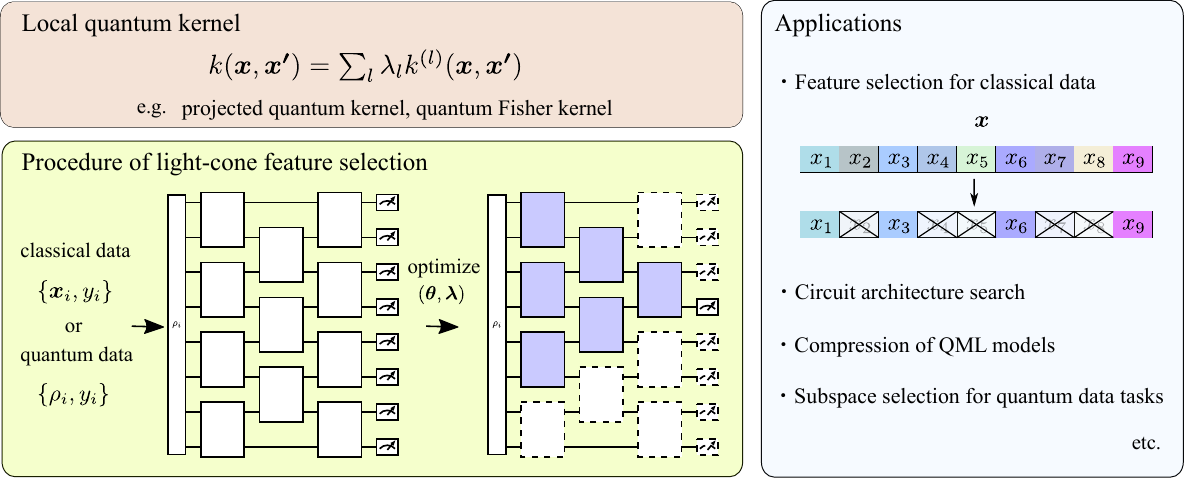}
    \caption{Summary of this work. Our proposed scheme selects relevant light-cone features for QML tasks through the training of parameters $(\bm{\theta},\bm{\lambda})$ in local quantum kernels (note that the local term $k^{(l)}(\bm{x},\bm{x'})$ includes tunable parameters $\bm{\theta}$).
    With the light-cone feature selection, we can handle not only the feature selection for classical data, but also the subspace selection for quantum data tasks, which conventional feature selection techniques cannot deal with.}
    \label{fig:summary}
  \end{center}
\end{figure*}

Quantum enhancement of machine learning attracts much attentions due to provable advantages of certain quantum machine learning (QML) methods over classical counterparts~\cite{wiebe2012quantum,aimeur2013quantum,rebentrost2014quantum,lloyd2014quantum,biamonte2017quantum}. 
Of particular interest are QML approaches for data analysis. 
A family of fruitful results have theoretically shown that tailored quantum models can efficiently learn patterns of synthetic datasets that are hard for classical ones to find~\cite{liu2021rigorous,jager2023universal,Muser2023tos}.
Therefore, the next step is to quest whether such advantages can be translated into practical machine learning tasks.

A primary candidate of supervised QML models in the near-term is the so-called quantum neural networks~\cite{schuld2014quest}.
The key idea of this approach is to utilize the Hilbert space as the feature space for machine learning tasks by exploiting the power of parameterized quantum circuits (PQCs) and data-embedding circuits~\cite{havlivcek2019supervised,schuld2019quantum}. 
Due to the mathematical equivalence, learning models based on quantum kernels can also be in the same class of models as the quantum neural networks~\cite{schuld2021supervised}; quantum neural networks and quantum kernel methods are categorized as explicit and implicit models within the same concept of QML, respectively~\cite{jerbi2023quantum,yano2021efficient,nghiem2021unified,herman2023expressivity,sweke2023potential}. 
Focus of this paper is on these supervised QML models.

Thus far, a number of attempts have been made to understand these quantum models analytically. 
In contrast to the original expectation, some literatures have elucidated that the high expressivity of the models causes trainability problems such as barren plateaus~\cite{mcclean2018barren,cerezo2021cost,holmes2021barren,holmes2022connecting,leone2022practical} and the exponential concentration~\cite{huang2021power,kubler2021inductive,thanasilp2022exponential,suzuki2022quantum}. 
This immediately suggests the need to carefully construct quantum models for real-world applications. 
A remedy for the issue in classical machine learning is feature selection~\cite{guyon2003introduction,saeys2007review,chandrashekar2014survey,pudjihartono2022review}. 
With the technique, redundant features can be removed and hence the predictive performance improves. 
Moreover, the model’s interpretability can be enhanced thanks to the reduced features.

We note that feature selection techniques have been studied in the quantum computing community. 
For example, Quadratic Unconstrained Binary Optimization problems corresponding to the classical feature selection have been introduced, so that quantum computing can deal with it to boost the performance~\cite{zoufal2023variational,mucke2023feature,hellstern2023quantum}. 
Other works use quantum support vector machines with multi-objective genetic algorithms~\cite{wang2023novel}, quantum circuit evolution algorithms~\cite{albino2023evolutionary} and graph-theoretic approach using quantum computers~\cite{chakraborty2020hybrid} for the performance improvement. 
However, the usability of classical feature selection could be limited for QML tasks.
While classical feature selection methods still play a critical role in embedding data on quantum circuits, it does not help understand why some data embedding quantum circuits work better than others. 
More importantly, when dealing with quantum data tasks such as quantum phase recognition, classical input features do not explicitly appear in input quantum data and thus one cannot employ the classical techniques straightforwardly.

In this work, we propose a feature selection method with a specific focus on QML problems. 
More specifically, we treat the light-cone structures (i.e., subspaces) of local quantum kernels as quantum features and then select a subset of beneficial ones through training of parameters in the kernels.
Note that, by local quantum kernels, we mean a family of quantum kernels that measure local similarity between a pair of data; examples are projected quantum kernels (PQKs)~\cite{huang2021power} and quantum Fisher kernels (QFKs)~\cite{suzuki2022quantum}.
Thanks to the extension, the proposed framework is applicable not only to the feature selection of classical inputs, but also to the quantum data tasks that conventional feature selection methods cannot handle.
To the best of our knowledge, this is the first work to introduce a QML-oriented feature selection method. 
Numerical simulations using toy tasks demonstrate the effectiveness of our scheme for four applications: (1) classical feature selection, (2) circuit architecture search for data embedding, (3) the compression of QML models and (4) relevant subspace selection for quantum data tasks.
These results indicate that our proposal will encourage practitioners to use QML models for real-world applications.
Figure~\ref{fig:summary} summarizes our work.
In addition, as elaborated in Conclusion \& Discussion section, this scheme is linked to the argument about classical simulability and trainability~\cite{cerezo2023does}.
Therefore, our scheme could work as a practical test to see if quantumness is really needed for the QML tasks at hand, while this is left for future work.

The rest of this paper is organized as follows. 
In Section~\ref{sec:proposal}, we provide the settings of QML framework and elaborate on our proposal.
Then, Section~\ref{sec:numerics}
demonstrates numerical simulations to see the effectiveness and versatility of our proposal.
Lastly, we discuss the potential of our methods in practical situations and then conclude this paper in Section~\ref{sec:conc}.

\section{Method} \label{sec:proposal}
\subsection{Framework of Quantum Machine Learning}
In this work, we focus on supervised quantum machine learning (QML) tasks; given a training dataset that consists of input data and the corresponding targets, our goal is to train a QML model so that its output accurately predicts the targets of new unseen data. 
Here, we consider the case where the dataset is either classical or quantum.

\begin{figure}[h]
  \begin{center}
    \includegraphics[scale=0.9]{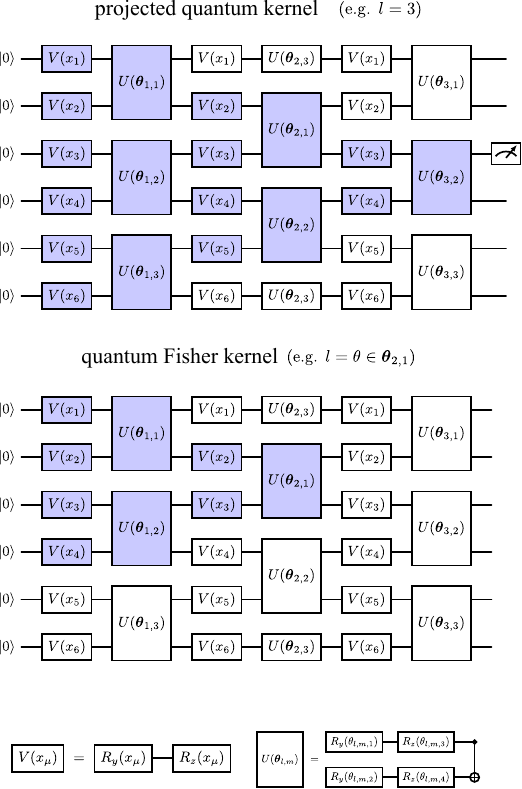}
    \caption{Examples of embedding quantum circuits with the depth $L=3$ and six features $\bm{x}=[x_1,\ldots,x_6]$ for the PQK and the QFK.
    The colored gates represent the light-cone of these quantum kernels; each shows the light-cone of measurement on $l$-th qubit with $l=3$ to construct
    the PQKs in Eq.~\eqref{eq:pqk}, and that of the gate including the $l$-th parameter $l=\theta\in\bm{\theta}_{2,1}$ for the QFK in Eq.~\eqref{eq:qfk}.}
    \label{fig:qc1}
  \end{center}
\end{figure}

In case of classical data, the dataset is composed of pairs of classical input $\bm{x}_i$ and its target value or label $y_i$; an example of the input is a vectorized $d$-dimensional image $\bm{x}\in\mathbb{R}^{d}$.
For QML models to deal with the classical data, we have to encode the input to quantum states via embedding quantum circuits. 
Namely, we apply an embedding circuit $V(\bm{x}_i)$ to an initial state $\rho_{0}$ to construct data-dependent quantum states $\rho_{i}=V(\bm{x}_{i})\rho_{0} V^\dagger(\bm{x}_{i})$. 
We can also utilize parameterized quantum circuits (PQCs) $U(\bm{\theta})$ with tunable parameters $\bm{\theta}$ to find better features in the space. 
Typically, we employ data re-uploading technique~\cite{perez2020data} to create quantum states from classical inputs;
\begin{equation}
 \rho_i(\bm{\theta})= U(\bm{x}_i,\bm{\theta})\rho_{0} U^\dagger(\bm{x}_i,\bm{\theta})
\end{equation}
with $L$-layer data re-uploading quantum circuits
\begin{equation}
    U(\bm{x},\bm{\theta})=\prod_{l=1}^{L} U(\bm{\theta}_{l}) V(\bm{x}).
\end{equation}
In the numerical experiments in Section~\ref{sec:numerics}, we use the re-uploading quantum circuits with a brick-like structure~\cite{cerezo2021cost}; an example for $n=6$ and $L=3$ is shown in Figure~\ref{fig:qc1}.
%%%%%%%%%% figure 2  %%%%%%%%%%%
%%%%%%%%%%%%%%%%%%%%%
As for quantum data, the dataset consists of quantum states and their targets, i.e., $\{\rho_i, y_i \}$~\cite{schatzki2021entangled,perrier2022qdataset}. 
In this case, we employ PQCs $U(\bm{\theta})$ to the quantum inputs to find better representation for the tasks;
\begin{equation}
 \rho_{i}(\bm{\theta }) = U(\bm{\theta})\rho_{i} U^{\dagger}(\bm{\theta}).
\end{equation}

Previous work defined two approaches of constructing quantum models~\cite{jerbi2023quantum}: explicit models where model predictions are given by measuring a certain Hermitian operator $O$ on quantum states, and implicit models where the inner products of quantum states called quantum kernels are used for predictions. 
Then, with a training dataset $S=\{\bm{x}_{i}, y_i\}_{i=1,\ldots,N}$ ($S=\{\rho_i, y_i\}_{i=1,\ldots,N}$) of size $N$, we perform optimization to obtain performant quantum models. 
For explicit models, we optimize $\bm{\theta}$ to minimize a loss function $\mathcal{L}(\bm{\theta})$ represented as
\begin{equation}
    \mathcal{L}(\bm{\theta}) = \sum_{i=1}^{N} l(f_i(\bm{\theta}),y_i)
\end{equation}
where $f_{i}(\bm{\theta})=\mathrm{Tr}[\rho_{i}O]$ is the output of explicit model and $l(f_{i}(\bm{\theta}),y_i)$ is the loss such as squared loss, i.e., $l(f_{i}(\bm{\theta}),y_i)=(f_{i}(\bm{\theta})-y_i)^2$.
On the other hand, the representer theorem~\cite{kimeldorf1971some,smola1998learning,hofmann2008kernel} guarantees that an optimal solution of implicit models can be in the form of $f_{i}^{opt} (\bm{x}) = \sum_i \alpha_{i}^{opt} k(\bm{x}_i, \bm{x})$ with a kernel function $k(\bm{x},\bm{x'})$.
A common choice of the kernel for QML is the fidelity quantum kernel~\cite{havlivcek2019supervised} expressed as 
\begin{equation} \label{eq:fid_qk}
    k_{Q}(\bm{x}_i, \bm{x}_j)= \mathrm{Tr}[\rho_i(\bm{\theta})\rho_j(\bm{\theta})].    
\end{equation}
For instance, the optimization problem for binary classification is reduced to minimizing the following function over the parameter $\bm{\alpha}$;
\begin{equation}
    \mathcal{L}(\bm{\alpha})=-\sum_{i} \alpha_i + \frac{1}{2}\sum_{i,j} \alpha_i\alpha_j y_iy_j k(x_i,x_j).
\end{equation}
Note that the kernel function can include the parameters $\bm{\theta}$ as shown in Eq.~\eqref{eq:fid_qk}; the parameters $\bm{\theta}$ can also be optimized separately~\cite{JMLR:v13:cortes12a} or simultaneously~\cite{Sydorov_2014_CVPR}. 
Basically, classical optimizers are employed for both models in the optimization.

\subsection{Light-Cone Feature Selection}
We here extend the idea of classical feature selection~\cite{guyon2003introduction} to propose its counterpart for QML.
The classical feature selection methods can be categorized into three types: the filter, wrapper, and embedded methods. 
The filter approach performs selections based on the relationship between features and the targets. 
The wrapper approach evaluates the performance of the model for different subsets of features and then select features based on a performance metric. 
Lastly, the embedded approach provides important features via training of models such as LASSO. 
See, e.g., Ref.~\cite{saeys2007review,chandrashekar2014survey,pudjihartono2022review} for more comprehensive review of classical feature selections. 

Our proposal is similar in spirit to the embedded approach; we select relevant local light-cone of quantum models through training of local quantum kernels. 
By local quantum kernels, we mean quantum kernels which measure local similarity between a pair of data. 
Note that fidelity quantum kernels in Eq.~\eqref{eq:fid_qk} measure global similarity and thus do not fall into the category.
The examples are projected quantum kernels (PQKs)~\cite{huang2021power} and simplified quantum Fisher kernels (QFKs)~\cite{suzuki2022quantum} defined as
\begin{equation}
\begin{split}
    k_{PQ}(\bm{x}_{i}, \bm{x}_{j}) &= \sum_{l} \lambda_{l}\mathrm{Tr}\Bigl[\rho_{i}^{(l)}(\bm{\theta}) \rho_{j}^{(l)}(\bm{\theta})\Bigr]
\end{split}
  \label{eq:pqk}
\end{equation}
and 
\begin{equation}
\begin{split}
    k_{QF}(\bm{x}_{i}, \bm{x}_{j}) &= \sum_{l} \lambda_{l}
  \mathrm{Tr}\Bigl[
    \rho_{0} \{ \tilde{B}_{i, \theta_{l}}, \tilde{B}_{j, \theta_{l}})\}
  \Bigr]
\end{split}
  \label{eq:qfk}
\end{equation}
respectively.
Here, $\rho_{i}^{(l)}=\mathrm{Tr}_{\bar{l}}[\rho_{i}(\bm{\theta})]$ is the partial trace of the quantum state $\rho_{i}(\bm{\theta})$ over all qubits except for the $l$-th qubit.
Also, 
$\tilde{B}_{i,\theta_{l}}=U_{1:l}^{\dagger}(\bm{x}_i,\bm{\theta})B_{\theta_{l}}U_{1:l}(\bm{x}_i,\bm{\theta})$ with  $B_{\bm{x}_i,\theta_l}=2i(\partial U(\bm{x}_i,\bm{\theta})/\partial \theta_{l})U^{\dagger}(\bm{x}_i,\bm{\theta})$, where $U_{k:l}(\bm{x},\bm{\theta})$ denotes a sequence of unitary gates from $U_{k}(\bm{x}_i,\theta_{k})$ to $U_{l}(\bm{x}_{i},\theta_{l})$ in the representation $U(\bm{x},\bm{\theta})=U_{N_{p}}(\bm{x},\theta_{N_p})\cdots U_{2}(\bm{x},\theta_{2}) U_{1}(\bm{x},\theta_{1})$ for the total number of parameters $N_p$.

These local quantum kernels can be expressed as summation of local terms, $k(\bm{x},\bm{x'})= \sum_l \lambda_l k^{(l)}(\bm{x},\bm{x'})$, where each local kernel $k^{(l)}(\bm{x},\bm{x'})$ has the corresponding light-cone, i.e, the subspace of quantum circuits (or quantum states) that are relevant to a certain operation like a local gate operation and measurement. 
For example, as illustrated in Figure~\ref{fig:qc1}, the light-cones obtained by measurement on $l$-th qubit ($l=3$) for the PQK and a gate operation including $l$-th parameter ($l=\theta\in\bm{\theta}_{2,1}$) for the QFK are represented by the colored gates.
We regard the light-cones induced by local quantum kernels as features and then select important ones through training of parameters $(\bm{\theta},\bm{\lambda})$ in quantum kernels. 

As for the training of quantum kernels, we maximize the kernel target alignment (KTA)~\cite{cristianini2001kernel} defined as
\begin{equation}
  \textrm{KTA} = \frac{\sum_{pq} y_{p} y_{q} k(\bm{x}_{p},\bm{x}_{q})}
  {\sqrt{\sum_{pq} k(\bm{x}_{p},\bm{x}_{q})^{2}} \sqrt{\sum_{pq}y_{p}y_{q}}}.
  \label{eq:kta}
\end{equation}
We note that the KTA is a common choice as an objective function for training kernels in both classical~\cite{JMLR:v13:cortes12a,cristianini2001kernel} and quantum cases~\cite{kubler2021inductive,thanasilp2022exponential,miyabe2023quantum}. 
In this work, the optimization of $(\bm{\theta},\bm{\lambda})$ is done in two steps: (1) with $\bm{\lambda}$ fixed, $\bm{\theta}$ are optimized via the Adam optimizer by computing the gradients of Eq.~\eqref{eq:kta} with respect to $\bm{\theta}$, and then, (2) $\bm{\lambda}$ are optimized following the approach proposed in Ref.~\cite{JMLR:v13:cortes12a}:
\begin{eqnarray}
  \bm{\lambda}^{\textrm{opt}} &=& \frac{\bm{v^{*}}}{|\bm{v^{*}}|_{1}},
  \label{eq:muopt}\\
  \bm{v}^{*} &=& \textrm{arg}\min_{\bm{v} \geq \bm{0}}
  \Bigl[\sum_{ij}v_{i}M_{ij}v_{j} - 2 \sum_{i}v_{i}a_{i}\Bigr],
  \label{eq:vopt}\\
  M_{ij} &=& \sum_{pq}k^{(i)}(\bm{x}_{p},\bm{x}_{q}) k^{(j)}(\bm{x}_{p},\bm{x}_{q}), \\
  a_{i} &=& \sum_{pq} k^{(i)}(\bm{x}_{p},\bm{x}_{q}) y_{p}y_{q}.
\end{eqnarray}
This two-step process is iteratively carried out until the value of the KTA stops changing significantly. 
The optimization can be done efficiently, as the first step relies on gradient-based algorithms and the solution of the second step can be obtained by solving a quadratic program. 
In our numerical experiments,  Eq.~\eqref{eq:vopt} in the second step is solved using CVXOPT library~\cite{andersen2011,cvxopt}.
After the optimization, we select relevant light-cones based on the amplitude of parameters $\bm{\lambda}$; that is, we interpret the local kernel $k^{(l)} (\bm{x},\bm{x'})$ whose parameter $\lambda_{l}$ has a large absolute value as important.

We remark that our proposal is more versatile than classical feature selection methods in QML tasks.
This scheme can be used not only for feature selection of classical features as the conventional methods do, but also for seeking relevant subspace of embedding quantum circuits.
More importantly, the proposed method is applicable to quantum data tasks where classical features do not appear and hence conventional ones cannot work.
In addition, while some previous works proposed methods for feature selection using quantum computing devices, all of them focus on improvement of the classical feature selection~\cite{zoufal2023variational,mucke2023feature,hellstern2023quantum,wang2023novel,chakraborty2020hybrid,albino2023evolutionary}.
Therefore, to the best of our knowledge, this work is the first to propose a QML-oriented feature selection scheme.

\section{Numerical Demonstration}  \label{sec:numerics}
In what follows, we perform numerical simulations to demonstrate efficacy and versatility of our proposal.
More specifically, we deal with four application tasks: (1) feature selection of classical inputs, (2) circuit architecture search for data embedding, (3) compression of QML models and (4) subspace feature selection for quantum data. 
For all the simulations shown below, PennyLane library~\cite{bergholm2022pennylane} is used for quantum circuit simulation. 
Also, in the learning tasks, we used 80 data points for both training and test, where models’ performance are checked for different 25 settings of data samples and initial parameters $\bm{\theta}$; five sets of data samples and five different parameters are prepared and then we tried all combinations of these settings.
As for local quantum kernels, we set a uniform weights to the initial parameters $\bm{\lambda}$, i.e., $\lambda_{l}=1/$(the number of terms in the local quantum kernel) for all $l$. 
The learning rate and the number of iterations in the Adam optimization step are set to 0.01 and 500, respectively.

\subsection{Feature Selection of Classical Inputs}
To begin with, we check the performance of our scheme for classical feature selection tasks. 
Here, we utilize the importance of light-cones determined by our proposal to seek out relevant classical features; that is, we evaluate the importance of the features based on a quantity determined by the optimized parameters $\bm{\lambda}$.
We call the quantity the importance score. 
In this study, we define the importance score $P(\mu)$ of a classical feature $\mu$ as  
\begin{equation} \label{eq:importance_score}
 P(\mu) = \frac{1}{\mathcal{N}}\sum_{l} w_{\lambda_{l}}(\mu) \lambda_{l} 
\end{equation}
where $w_{\lambda_{l}}(\mu)$ is the frequency for the classical feature $\mu$ to appear in the $l$-th local quantum kernel $k^{(l)}(\bm{x},\bm{x’})$, and $\mathcal{N}=\sum_{\mu} \tilde{P}(\mu)$ with $\tilde{P}(\mu)=\sum_{l} w_{\lambda_{l}}(\mu) \lambda_{l}$ is the normalization constant.
Recall that $\bm{\lambda}$ satisfy $\sum_l \lambda_l=1$ with $\lambda_l \ge 0$. 
In case of the light-cone structure of the PQK shown in Figure~\ref{fig:qc1} (i.e., $\lambda_l$ with $l=3$ depicted in the upper quantum circuit), $ w_{\lambda_{l}}(\mu) = 1$ for $\mu = x_1,x_6$, $ w_{\lambda_{l}}(\mu) = 2$ for $\mu = x_2,x_5$, and $ w_{\lambda_{l}}(\mu) = 3$ for $\mu = x_3,x_4$. 
We notice that one can take another quantity as the importance score, so long as it tells the relevance of features in the tasks.

As a toy example, we consider a so-called parity-based binary classification task introduced in Ref.~\cite{jerbi2023quantum}. 
The goal of this task is to classify $d$-dimensional binary input $\bm{x}\in \{-1,1\}^d$ whose label is determined based on its parity for a certain subset $A_{p}\subset \{1,\ldots,d\}$; namely, the input’s label is assigned according to $y(\bm{x})=\prod_{i\in A_{p}} x_i$. 
We chose this task because important classical features are clearly $A_{p}$ and hence we can interpret the performance of our proposal easily.  
In the following, we consider a dataset with $A_{p}=\{1,2\}$, which we denote as parity(1,2).
Here, the number of qubits $n$ is equal to the dimension of inputs $d$ and we set $d=8$.
Also, the data re-uploading quantum circuits with the depth $L=3$ is used.

\begin{figure}
  \begin{center}
    \includegraphics{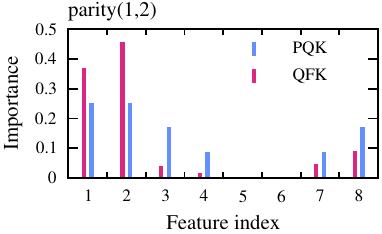}
    \caption{Importance scores of features for the parity(1,2) dataset.}
    \label{fig:parity_importance_nq08}
  \end{center}
\end{figure}

Figure~\ref{fig:parity_importance_nq08} shows the importance scores of the local quantum kernels for the parity(1,2) dataset.
Features whose importance scores are higher than the threshold $P_{th}=0.1$ for QFKs are $x_1$ and $x_2$, suggesting our scheme with QFKs can pinpoint key features for the task.
As for PQKs, features $x_1$, $x_2$, $x_3$ and $x_8$ witness the scores larger than $P_{th}$.
Hence, the PQKs can also select relevant features on the dataset.
The features chosen by the PQKs include more features than the minimum required.
This is because the number of classical features in the light-cone induced by a local measurement gets larger as the depth increases~\cite{suzuki2023effect}.
We note that the light-cone of $l$-th term of QFKs is determined by the position of gates with the local parameter $\theta_l$~\cite{suzuki2022quantum}, and thus only necessary features can be selected in this case.
Actually, the performance difference between these two kernels can be seen by looking at the KTAs; the values of the KTAs for PQKs and QFKs are 0.605 and 0.987, respectively.
Recall that the KTA of the best kernel outputs one, and the larger value of the quantity indicates the better performance. 
Hence, the performance gap between the PQK and the QFK would be enlarged as the depth increases. 
Yet, we could resolve the problem in the PQKs, e.g., by using different feature scores.

\subsection{Circuit Architecture Search for Data Embedding}
Next, we check if our scheme can be exploited to find better embedding quantum circuits.
For classical data tasks, embedding quantum circuits play a crucial role in the performance, since the data distribution in the Hilbert space is determined by them.
Previous works introduced techniques to increase the expressivity in terms of Fourier analysis~\cite{schuld2021effect,shin2023exponential}. 
Still, there is room for investigation in guidelines for constructing performant embedding quantum circuits.
Hence, we here propose a new embedding strategy based on our light-cone feature selection technique.

Specifically, this work focuses on the order in which classical features are encoded onto qubits. 
A straightforward and the most popular strategy is to assign a feature $x_i$ to the $i$-th qubit for $i=1,\ldots,n$.
However, due to the restricted connectivity of entangling gates, underlying patterns of data in the Hilbert space might not be found with the standard encoding approach in some cases; for example, the features assigned in the farthest qubits in the quantum circuit could be less correlated, leading to poor expressivity of the model for certain tasks. 
With this in mind, we introduce a problem-specific ordering of data encoding. 
Here, we consider the following scheme: we start with the standard encoding described above and, after the light-cone feature selection optimization, the order of features encoded onto the qubits are sorted in descending order of their importance scores in Eq.~\eqref{eq:importance_score}.
Let us note that our technique is applicable to more complicated settings; one could also use this idea to find a better gate set among some candidates.

\begin{figure}
  \begin{center}
    \includegraphics{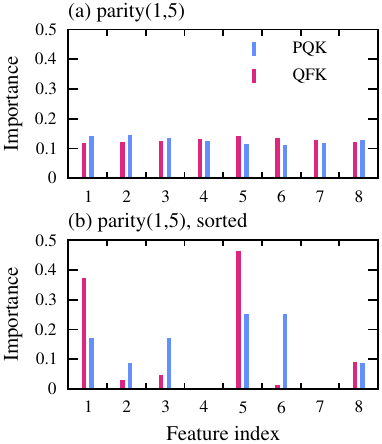}
    \caption{Importance scores of features for the parity(1,5) datasets (a) without and (b) with re-ordering of input encoding.}
    \label{fig:parity_importance_nq08_15}
  \end{center}
\end{figure}

For the proof-of-concept demonstration of this scheme, we again work on the parity-based binary classification task. 
In the numerical experiments, we consider the dataset with $A_{p}=\{1,5\}$ denoted as parity(1,5). 
Table~\ref{tbl:parity_kta} shows the KTA values for the cases with and without the re-ordering technique. 
Here, the optimization of local kernels are done three times for both cases.
It turns out that the KTAs increase from -0.033 to 0.987 (-0.015 to 0.560) for QFKs (PQKs). 
The performance improvement can be interpreted in terms of feature selection. 
Figure~\ref{fig:parity_importance_nq08_15} shows the importance scores for these cases. 
When the technique is not employed, every feature is evenly regarded as important for QFKs and PQKs.
Together with the KTA values, this indicates that the optimization works poorly because the restriction on entangling gates limits the spread of light-cones.
Note that a few local terms in PQKs can cover features $x_1$ and $x_5$, but they are so far away that optimization could not lead to good prediction. 
On the other hand, QFKs and PQKs can successfully select the relevant features after the re-ordering, suggesting that this technique facilitates the optimization and results in the better performance.

\begin{table}

\caption{The KTA values for the parity(1,5) and the breast-cancer datasets.}
\centering
  \begin{tabular}{llccc}
    \hline
    parity(1,5) & &  PQK & QFK \\
    \hline \hline
    without sorting & train  & 0.037 & 0.077 \\
                & test &   -0.015 & -0.033 \\ \hline
    with sorting & train &       0.562 & 0.987  \\
           & test        & 0.560 & 0.987  \\
    \hline
    \rule{0pt}{4ex} \\
    \hline
    breast-cancer & & PQK  & QFK  \\
    \hline \hline
    without sorting & train  & 0.397 & 0.749 \\
                & test &   0.366 & 0.719 \\ \hline
    with sorting & train &       0.389 & 0.752  \\
           & test        & 0.377 & 0.726  \\
    \hline
    
  \end{tabular}
  \label{tbl:parity_kta}
\end{table}

\begin{figure*}[t]
  \begin{center}
    \includegraphics{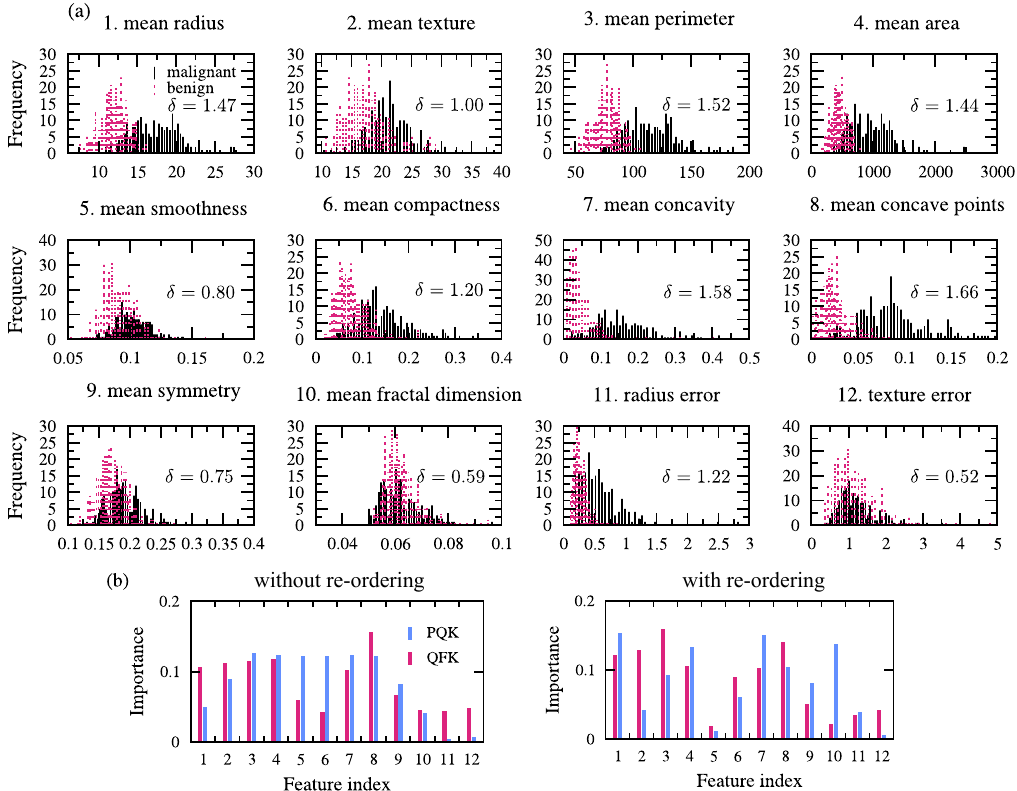}
    \caption{Results on the breast-cancer dataset. (a) Histograms of the 12 features of the breast-cancer dataset used in this work are illustrated. In each panel, $\delta$ denotes the L1 norm of the distributions  computed with the fixed number of bins (100).
      (b) Importance scores are calculated for the cases with (right) and without (left) re-ordering technique.
    }
    \label{fig:bc_hist_imp}
  \end{center}
\end{figure*}

Moreover, we work on the benchmark dataset, the breast-cancer dataset, to see the performance for a more practical task.
This is a classical dataset for binary classification where 212 data points are labeled as ``malignant" and 357 as ``benign".
In the numerical experiments, we choose $d$($=12$) features out of 30 and then rescale the data so that it has a mean zero and unit variance.
The distribution of each feature is shown in Figure~\ref{fig:bc_hist_imp}~(a), in which we show the L1 norm of ``malignant" and ``benign" distributions $\delta$ to show the discrepancy (i.e., separability) of them.
The dataset is taken from scikit-learn~\cite{scikit-learn}.
Also, we employ the same quantum circuits as the parity-based classification tasks.

As shown in Table~\ref{tbl:parity_kta}, the re-ordering technique increases the KTA values of QFKs (PQKs) from 0.719 to 0.726 (from 0.366 to 0.377) for test, while the KTA of PQKs for training slightly decreases.
The improvement in the test data is reasonable because more compact models are constructed by our scheme. 
Figure~\ref{fig:bc_hist_imp}~(b) illustrates importance scores with and without re-ordering.
We observed that four features ($x_{5},x_{10}, x_{11}, x_{12}$) for QFKs with re-ordering show smaller importance scores than the smallest value for the case without the technique (0.04 for $x_{6}$), implying that the model gets sparse thanks to our scheme.
Similarly, the number of important features selected by PQKs with the threshold $P_{th}=0.1$ is reduced by this scheme; six features ($x_{3}$, $x_{4}$, $x_{5}$, $x_{6}$, $x_{7}$ and $x_{8}$) are higher than the threshold for the case without the method, whereas five ($x_1$, $x_4$, $x_7$, $x_8$ and $x_{10}$) are selected for re-ordering. 
From these results, we can confirm the effectiveness of the new encoding scheme based on the importance score.

\subsection{Compression of Quantum Machine Learning Models}
Another application where our proposal can be utilized is the compression of quantum circuit models. 
The model compression aims to reduce the size of machine learning models, e.g., the number of parameters in neural networks~\cite{cheng2017survey}; as a result, resource efficiency, faster inference and better generalization can be realized. 

Especially, pruning is a powerful tool to reduce the model size in classical machine learning. 
With the pruning technique, one can construct a compressed model that requires less resources to execute inference and is comparable to the original model in the performance. 
For example, parameters of neural networks are pruned based on their magnitudes~\cite{frankle2018lottery} (i.e., the parameters with small values are removed). 
Indeed, the magnitude-based pruning method has been explored in variational quantum algorithms~\cite{sim2021adaptive,wang2022quantumnas}. 
On the other hand, careful consideration is needed when applying the magnitude-based pruning to data re-uploading quantum circuits.
More concretely, this technique fails to prune the embedding layer, suggesting its inability to reduce the model width (i.e., the number of qubits). 
This also means that the reduction in the depth might not be huge as the embedding layers will never be removed.

With a focus on the problem, we propose a new pruning method using the light-cone feature selection. 
Here, we evaluate the magnitudes of parameters $\bm{\lambda}$ to determine which part of quantum circuits are redundant. 
Recall that $\bm{\lambda}$ are the weights for local terms and thus their amplitudes tell the importance of each light-cone. 
Consequently, this scheme can prune even the embedding layers.

\begin{figure}[t]
  \begin{center}
    \includegraphics[scale=0.6]{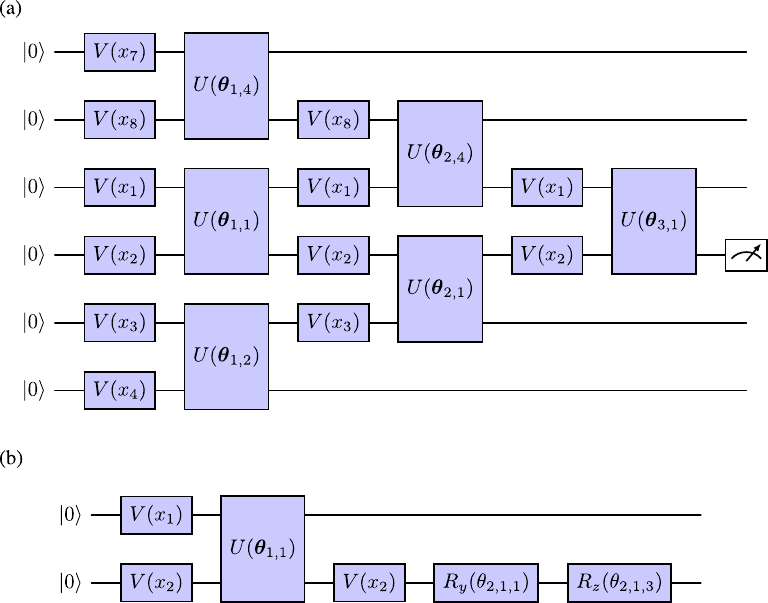}
    \caption{Pruned quantum circuits. Here, we consider the light-cone of the local term of (a) the PQK and (b) the QFK whose parameter $\lambda$ is largest for the parity(1,2) dataset.}
    \label{fig:qc_qfklc}
  \end{center}
\end{figure}

To see how the pruning scheme works, we consider the parity(1,2) dataset. 
To be precise, we prune all the light-cones of models except for the one with the largest parameter $\lambda$ and check the relationship to the performance of the corresponding compressed model. 
Figure~\ref{fig:qc_qfklc} depicts the quantum circuits constructed by removing all but the light-cone associated with the largest $\lambda$ for the PQK and QFK, respectively. 
The light-cone chosen by this scheme is reasonable as the important subspace in this task is the part of the circuit where features $x_1$ and $x_2$ are included.
Moreover, the performances of the pruned models are comparable to the original models; the KTA value of PQKs (QFKs) just decreases from 0.605 to 0.600 (0.987 to 0.983) by pruning.
Thus, our numerical experiment suggests the validity of our proposal. 

\subsection{Subspace Feature Selection for Quantum Data}

Lastly, we work on a task in which the goal is to find relevant light-cones (i.e., subspaces) of quantum data. 
We recall that conventional feature selection methods cannot deal with the tasks because classical features do not appear. 
The advantage of the approach would be that one can construct an efficient and better-performing QML model by removing irrelevant subspaces of quantum states.

\begin{figure*}
  \begin{center}
    \includegraphics{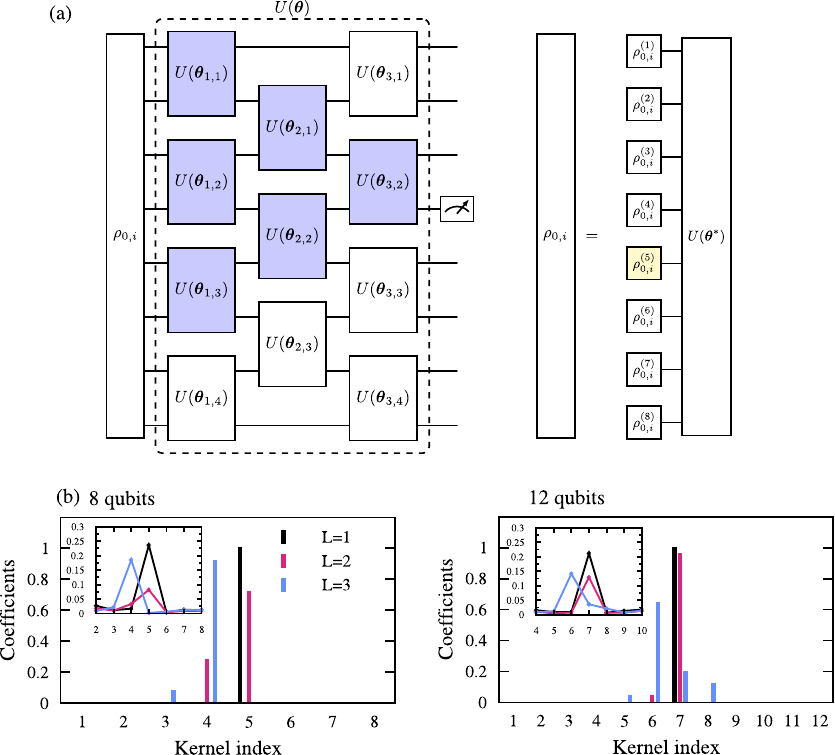}
    \caption{Results for a toy quantum data task. (a) The setting of the QML model with the PQC's depth $L=3$ and the quantum data considered in this experiment are illustrated. The colored quantum state $\rho_{0,i}^{(l)}$ with $l=n/2+1$ (in this case, $l=5$) is a key state used to assign its label $y_i$. (b) Optimized parameters $\bm{\lambda}$ of the PQKs for quantum data tasks with $n=8,12$ are shown. Here, quantum circuits with $L=1,2,3$ are used. The insets of figures illustrate the averaged KTA of PQKs calculated using a single reduced density matrix $\rho_{i}^{(l)}(\bm{\theta})$; that is, $l$ in the horizontal axis denotes the PQK in Eq.~\eqref{eq:pqk} with $\lambda_{l}=1$ and 0 otherwise.
    }
    \label{fig:qdata_coeffs}
  \end{center}
\end{figure*}

To see the effectiveness of our proposal, we consider a toy binary classification task with a synthetic quantum dataset $\{\rho_{i}, y_i\}$. 
Here, the input quantum state is represented as $\rho_{i}=U(\bm{\theta}^{*})\rho_{0,i}U^\dagger(\bm{\theta}^{*})$, where $\rho_{0,i}=\rho_{0,i}^{(1)}\otimes\rho_{0,i}^{(2)}\otimes \ldots \otimes \rho_{0,i}^{(n)}$ is the tensor product of single-qubit Haar random states $\{\rho_{0,i}^{(l)}\}_{l=1,\ldots,n}$ and $ U(\bm{\theta}^{*})$ denotes the unitary operator with certain fixed parameters $\bm{\theta}^{*}$. 
Then, the label is determined by $y_i=\text{sign}(\mathrm{Tr}[Z_{m} \rho_{0,i}])$ with the Pauli $Z$ operator acting on the $m$-th qubit $Z_{m}$; see Figure~\ref{fig:qdata_coeffs}~(a) for the details of this numerical experiment.
We construct the dataset inspired by actual quantum data tasks; like the quantum phase recognition tasks in Ref.~\cite{lake2022exact}, we assume that there exist a unitary operator that can transform input quantum state $\rho_{i}$ to another state in the form of $\tilde{\rho}_{i,r}\otimes \rho_{irr}$ with a relevant quantum state $\tilde{\rho}_{i,r}$ and an irrelevant state $\rho_{irr}$.
By a relevant quantum state, we mean that a simple measurement on that state is enough to identify the label or the property of the original state.
In this work, we focus on PQKs using alternating layered ansatzes~\cite{cerezo2021cost} with the number of qubit $n=8,12$ and the circuit depth $L=1,2,3$. 
Note that we do not take into account QFKs since the kernels are not defined via input quantum states $\rho_i$, but via the unitary representation of data $U(\bm{x_i},\bm{\theta})$. 
Also, we set $m= n/2 +1 $.

Figure~\ref{fig:qdata_coeffs}~(b) shows the amplitudes of $\bm{\lambda}$ after optimization of $(\bm{\theta},\bm{\lambda})$ for each setting. 
We clearly see that PQKs for the depth $L=1,2$ can point out the relevant quantum subspace. 
For $L=3$, however, this method selects the $n/2$-th qubit as the most important. 
This would be attributed to how the parameters $\bm{\theta}$ are optimized; the light-cone induced by the measurement of $n/2$-th qubit also covers the key qubit and the optimization in the light-cone is better than others.
To see if the understanding is correct, we compute the KTA values of all the possible PQKs where only a single-qubit reduced density matrix is used.
As shown in the insets of Figure~\ref{fig:qdata_coeffs}~(b), we found that the PQK with $\lambda_{n/2}=1$ and $0$  otherwise outputs the best KTA for $L=3$, indicating the reduced density matrix $\rho_{i}^{(n/2)}$ is the most relevant subspace.
In summary, we can numerically validate the potential of our scheme using a toy quantum data task.

\section{Conclusion \& Discussion} \label{sec:conc}

In this work, we propose a QML-oriented feature selection method where important light-cone features (subspaces) of quantum models are selected through training of local quantum kernels such as PQKs and QFKs.
Numerical simulations demonstrate that our scheme works well in (1) classical feature selection, (2) circuit architecture search for data embedding, (3) model compression and (4) subspace feature selection for quantum data.
These results suggest its versatility and encourage practitioners to use this technique for the application of QML models to real-world tasks.

As described in Section~\ref{sec:proposal}, our framework relies on the training of parameters $(\bm{\theta},\bm{\lambda})$ in quantum kernels; this means that the classical computational cost for the quantum kernel estimation scales at least quadratically with respect to the number of training data points.
Hence, there is room for investigation in improving its scalability. 
As the remedy to the similar issue in quantum kernel methods, previous works used techniques such as classical shadow and surrogate models based on random Fourier features~\cite{huang2021power,nakaji2022deterministic,shin2023analyzing}. 
Thus, it would be interesting to explore the improvement of its scalability in this direction.

According to Ref.~\cite{cerezo2023does}, the absence of trainability issues such as barren plateaus and exponential concentrations is strongly tied with classical simulability.
This indicates that, for classical data tasks, our technique might not work in the classically-hard regime; even if it works, the compressed model obtained through our method could be classically simulable. 
On the other hand, this also means that our method can be used as a practical test to see if the QML tasks really require quantumness. 
More significantly, our proposal is also applicable to quantum data tasks such as quantum phase recognition~\cite{cong2019quantum,lake2022exact,wu2023quantum} and entanglement detection~\cite{guhne2009entanglement} (i.e, quantum physics-related learning tasks). 
Therefore, our proposal could also be used for constructing efficient settings of such quantum tasks and the investigation of practical quantum advantages.

\begin{acknowledgements}
This work was supported by MEXT Quantum Leap Flagship Program Grant Number JPMXS0118067285 and JPMXS0120319794.
Y.S. was supported by Grant-in-Aid for JSPS Fellows 22KJ2709. 
\end{acknowledgements}

\bibliographystyle{unsrtnat}
\bibliography{reference}

\end{document}